\documentclass[a4paper,10pt]{article}

\usepackage[a4paper, total={6.5in, 8.5in}]{geometry}
\usepackage[utf8]{inputenc}
 \usepackage{graphics}
 \usepackage{graphicx}
 \usepackage{setspace}
\usepackage{amssymb}
\usepackage{amsmath}
\usepackage{amsthm}

\title{On the magneto-transport properties in the magnetic phase of BaFe$_{2-x}$TM$_x$As$_2$ (TM = Co, Ni): a magnetic excitations approach}
\author{J. P. Pe\~na$^1$, M. M. Piva$^2$,  P. F. S. Rosa$^{2,3}$, P. G. Pagliuso$^2$, C. Adriano$^2$,
T. Grant$^3$,\\ Z. Fisk$^3$,  E. Baggio-Saitovitch$^4$, P. Pureur$^1$}

\begin{document}
\begin{doublespace}

\maketitle

\begin{center}
 
$^1$Instituto de F\'isica, Universidade Federal do Rio Grande do Sul, Av. Bento Gon\c{c}alves 9500,  ZIP 15051, 91501-970, 
Porto Alegre, RS, Brazil\\

$^2$Instituto de F\'isica Gleb Wataghin, Universidade Estadual de Campinas, Rua Sérgio Buarque de  Holanda 777, ZIP 13083-970, 
Campinas, SP, Brazil\\

$^3$Department of Physics and Astronomy, School of Physical Sciences, University of California, 2186 Frederick Reines Hall, 
 ZIP 92697-4574, Irvine, CA,  USA
 
$^4$Centro Brasileiro de Pesquisas F\'isicas, Rua Dr. Xavier Sigaud 150, ZIP 22290-180, Rio de Janeiro, RJ, Brazil
 
\end{center}

\begin{abstract}

Because of their complex Fermi surfaces, the identification of  the physical phenomena 
contributing to electronic scattering in the Fe-based superconductors (FeSC) is a difficult task. 
Here, we report on the electrical resistivity, magnetoresistance (MR) and Hall effect in two series of BaFe$_{2-x}$TM$_x$As$_2$  \mbox{(TM = Co, Ni)} 
crystals with different values of $x$. 
The TM contents were chosen so that the majority of the investigated samples present an intermediate magnetically 
ordered state and a superconducting ground state.
We interpret the obtained results in terms of scattering of
charge carriers by magnetic excitations instead of describing them as resulting  
uniquely from effects related to multiple-band conduction. 
Our samples are single-crystals from the structural point of view and their overall magneto-transport properties
are dominated by a single magnetic state. 


\end{abstract}


\section{Introduction}

Electronic transport measurements are useful to survey the origin of electronic scattering  and the  excitation spectrum
of a material's superficial and bulk states.
However, in many cases several physical phenomena contribute to scattering events, so that the achievement
of a correct and unique interpretation of the
experimental results can become a major difficulty.
The electrical magneto-transport properties of the Fe-based superconductors (FeSC) are representative of these complex cases.
In most of these materials, the five different hole and electron bands forming the Fermi surface ~\cite{paglione, kishida},
and the exotic nematic/magnetic ordering established at the spin density wave ($SDW$) critical temperature, $T_{SDW}$,
make difficult to  establish what are the main factors ruling the charge transport in these compounds.

In the parent compound of the 122-family, BaFe$_2$As$_2$, both the transversal 
magnetoresistance (MR in the geometry $B\parallel c-$axis; $B\perp I $) and  the Hall coefficient,
$R_H$, exhibit unusual magnetic-field and temperature dependencies in the region $T < T_{SDW}$ ~\cite{hsueh, kkhuynh, albenque_lett, fang, pena}. 
In particular, the absolute values of the MR and $R_H$ sharply  increase when the temperature decreases below $T_{SDW}$.
These effects are commonly explained in the context of multiple band conduction models ~\cite{albenque_lett, fang, myi}. 
Such models require variations in the number and  mobility of the charge carriers all along the $SDW$ phase. 
Those variations are partially justified by the expected opening of a $SDW$-type gap.   
In fact, infrared and optical spectroscopy measurements show  that a spectral weight lost appears in the
low energy spectrum of BaFe$_2$As$_2$ below $T=T_{SDW}$ ~\cite{marsik2013, yin2}.
Nonetheless, Angle Resolved Photoemission Spectroscopy (ARPES) results do not support the opening of an actual gap  in the Fermi surface ~\cite{gliu2009}.
Instead,  a severe reorganization of the bands and the emergence of singular strong Fermi spots  are observed ~\cite{gliu2009}.
This reorganization of the bands provides additional channels for interband transitions  which could offer an interpretation of
the optical results different than that based on the opening of a $SDW$ gap  ~\cite{gliu2009}.
Additional ARPES results in Ref. ~\cite{shimojima2010} show that the electronic reconstruction below $T_{SDW}$ is highly orbital-dependent.
That characteristic makes the Fe$-3d_{xz}$ band dominant over the other iron bands in the magnetic phase ~\cite{shimojima2010},
contrasting with the more multi-orbital character of the PM phase. 
In conclusion,  ARPES results seem to remove some of the importance given to the gap opening and 
multiple band character to explain the transport phenomena in the $SDW$ phase of the FeSC.

Concerning the magnetic ordering, the multiple-band models consider that the Fe-based pnictides 
are well described by a fully itinerant picture where the opening of a normal $SDW$ gap plays the major role.
However, many works point out that both itinerant and localized nature of the magnetic moments have to be 
taken into account to properly describe the magnetic, transport and spectroscopic properties of these compounds
 ~\cite{yin2, gliu2009, zhao, liu, yin2011}.
In fact, by analyzing ARPES results, authors in Ref. ~\cite{yang2009} affirm that one could visualize the magnetic 
phase in BaFe$_2$As$_2$ within the perspective of effective local moments, so that the collinear $SDW$ order is
caused by the exchange interactions between the nearest neighbors and the next-nearest neighbors. 
Within this scenario, the $SDW$ naturally becomes commensurate in the parent compound without requiring 
the opening of a large gap on the Fermi surface or nesting of the hole and electron bands  ~\cite{yang2009}.
In consequence, evoking the partially local character of magnetic moments and the fact that 
magnetic excitations can strongly modify a material's quasi-particle scattering spectrum, 
some authors proposed that 
carrier scattering by magnetic excitations plays an additional, relevant mechanism for completely 
describing the magnetotransport properties of the Fe-based pnictides, including the TM-substituted systems \cite{pena, eom}.

Here we analyze the magnetotransport properties of two series of  BaFe$_{2-x}$TM$_x$As$_2$  samples, with \mbox{TM = Co} and Ni.
The TM content was varied within a range where the $SDW$ ordering is preserved in all studied crystals. Most of these samples
have a superconducting ground state, which is not necessarily associated with the same phase provoking the magnetically ordered phase. 
We show that the exotic behavior of the magneto-transport properties exhibited by the BaFe$_2$As$_2$ are displaced
to lower temperatures in the TM-substituted samples. 
Even so, the striking differences between these 
properties in the PM and $SDW$ phases remain while a magnetically ordered phase is manifested.
We argue that the description of the essential changes observed in the magneto-transport properties of the studied 
samples in the magnetically ordered state are mostly due to carrier scattering by magnetic excitations, although the 
effects from the multiple band character of the charge carriers and the Fermi surface reconstruction phenomenon can not be ignored.

\section{Experimental details}

Crystals of BaFe$_{2-x}$TM$_x$As$_2$ (TM = Co, Ni)
were synthesized by the auto-flux or by the In-flux methods.
The complete synthesis processes are reported in \mbox{Refs. \cite{garitezi, ychen}}.
Energy Dispersive Spectroscopy (EDS), X-Rays diffraction (XRD), and resistivity vs. temperature (four-probe-method)
measurements were performed in all samples for characterization.
We were not able to calculate the $x$ value from the EDS results in all cases. Then, the content of Co or Ni atoms was estimated by  a
comparison of the transition temperatures (magnetic and superconducting)  identified in the electrical resistivity curves with phase diagrams in 
literature \cite{ychen, canfield1, canfield2}. 
From XRD patterns we could identify only one crystalline phase from the recorded ($00l$) peaks of the Cu$k_{\alpha1}$ and Cu$k_{\alpha2}$ 
diffractions. 
Despite the single-crystal character of the samples,
the zero-field electrical resistivity curves revealed in some cases more than one magnetic or 
superconducting critical temperatures. 
We attribute these features to the presence of minoritary phases with different contents of the TM atoms
or partial superficial oxidation.
This effect is probably present, but went unnoticed, in several
samples of the 122-family of the Fe-based superconductors presenting simultaneously magnetic and superconducting states. 
Nevertheless, the highest magnetic transition temperature, labeled as $T_{SDW}$ and associated with a remarkable resistivity 
anomaly, characterizes the largely dominant electronic phase in all cases. 
Table \ref{resum} summarizes the parameters characterizing the main phase of the 10 studied samples 
(in five of these samples  TM = Co, and in the other five TM = Ni). 
These parameters are the $c-$axis lattice length, the TM content of the dominant electronic phase, and the respective
ordering temperature $T_{SDW}$.


The electrical transport measurements were carried out with the four-probe method 
in a low-frequency AC bridge of a commercial Quantum Design PPMS@ platform. 
Two contacts pads for current were attached with silver epoxy to the extremities of samples having the approximate form of parallelepipeds.
Two contact leads were attached to the same (opposite) edge of the sample for measuring the longitudinal (transversal) voltage.
Magnetic fields with magnitude in the range $B = 0$ and $B = \pm9$ T were applied parallel to the $c-$axis
and perpendicularly to the current. 
The planar component of the magnetoresistance  and the Hall resistivity were determined from 
the averages $\rho_{even}=\frac{\rho_++\rho_-}{2}$ and $\rho_{odd}=\frac{\rho_+-\rho_-}{2}$ of the longitudinal and transversal measurements,
respectivelly.
The term $\rho_{+/-}$ refers to the resistivity measured when the direction of the magnetic field was positive/negative
with respect to the vertical axis.


\begin{table}
\caption{Characterization parameters of the  main electronic phase of the samples studied here. 
NM is acronym of ``Not Measured''.
}\label{resum}
\small
\centering
\begin{tabular}{cccc|cccc}
 & \textbf{TM = Co} & & & &  \textbf{TM = Ni} & & \\
\hline
Sample & $c$ (\AA)& $x$ & $T_{SDW}$ (K) & Sample & $c$ (\AA)& $x$ & $T_{SDW}$ (K)  \\
\hline
Co-A & 13.028(2) & 0.023(2) & 119(1) & Ni-A & 13.058(2) &  0.015(2) & 121(1)  \\
\hline
Co-B & 13.003(2) & 0.032(3) & 115(2) & Ni-B & 13.018(2) &  0.030(2) & 105(2)  \\
\hline
Co-C & 13.013(2) & 0.037(4) & 113(4) & Ni-C & 13.038(2) &  0.033(2) & 101(3) \\
\hline
Co-D & 13.034(2) & 0.043(4) & 108(2) & Ni-D & 13.011(2) &  0.035(2) & 98(3) \\
\hline 
Co-E & 12.979(2) & 0.118(2) & 70(2)  & Ni-E & NM &  0.051(4) & 67(3) \\

\end{tabular}
\end{table}

\section{Results and discussion}

\subsection{Resistivity}

\begin{figure}
 \centering
\includegraphics[keepaspectratio,width =7.2truecm]{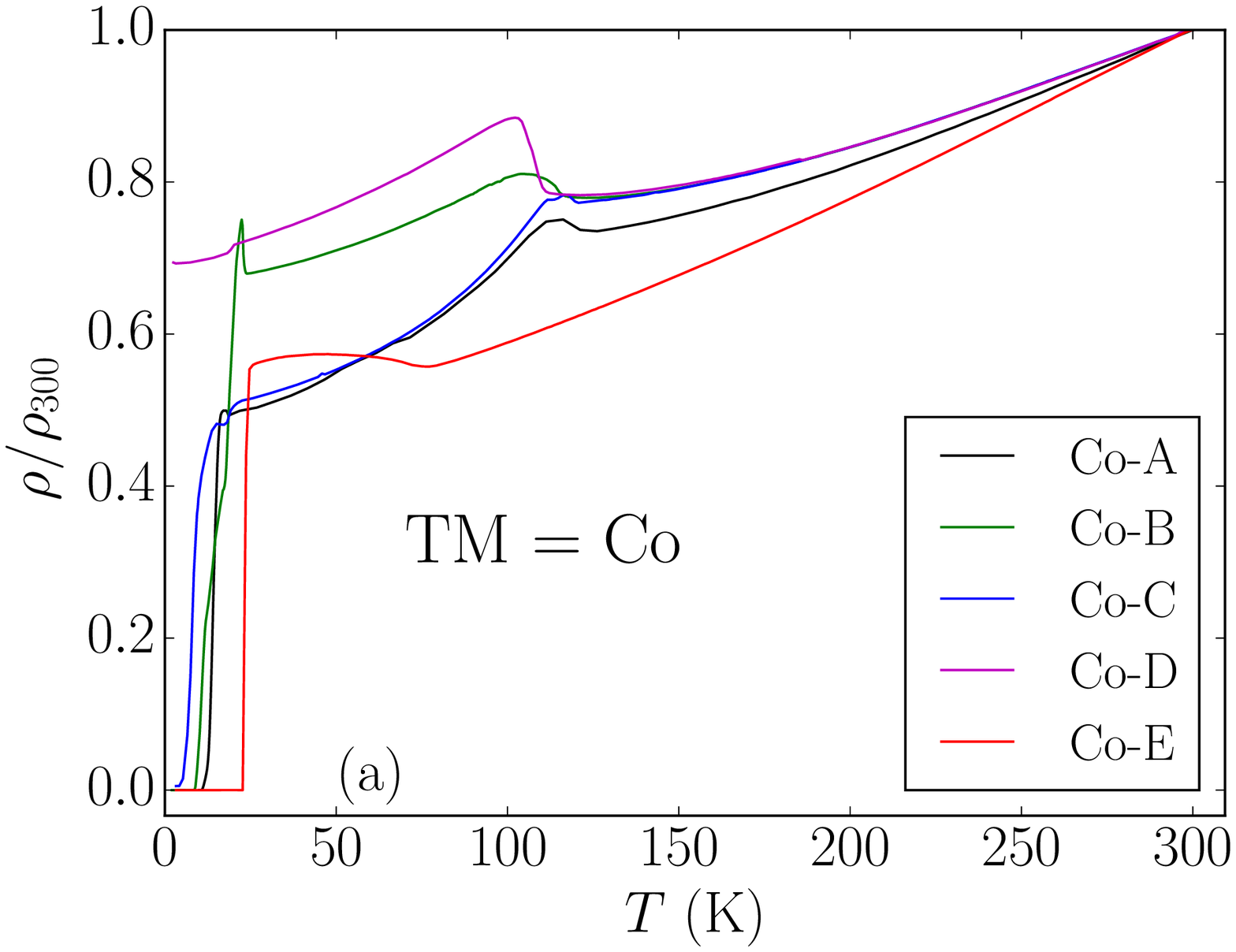}
\includegraphics[keepaspectratio,width =7.2truecm]{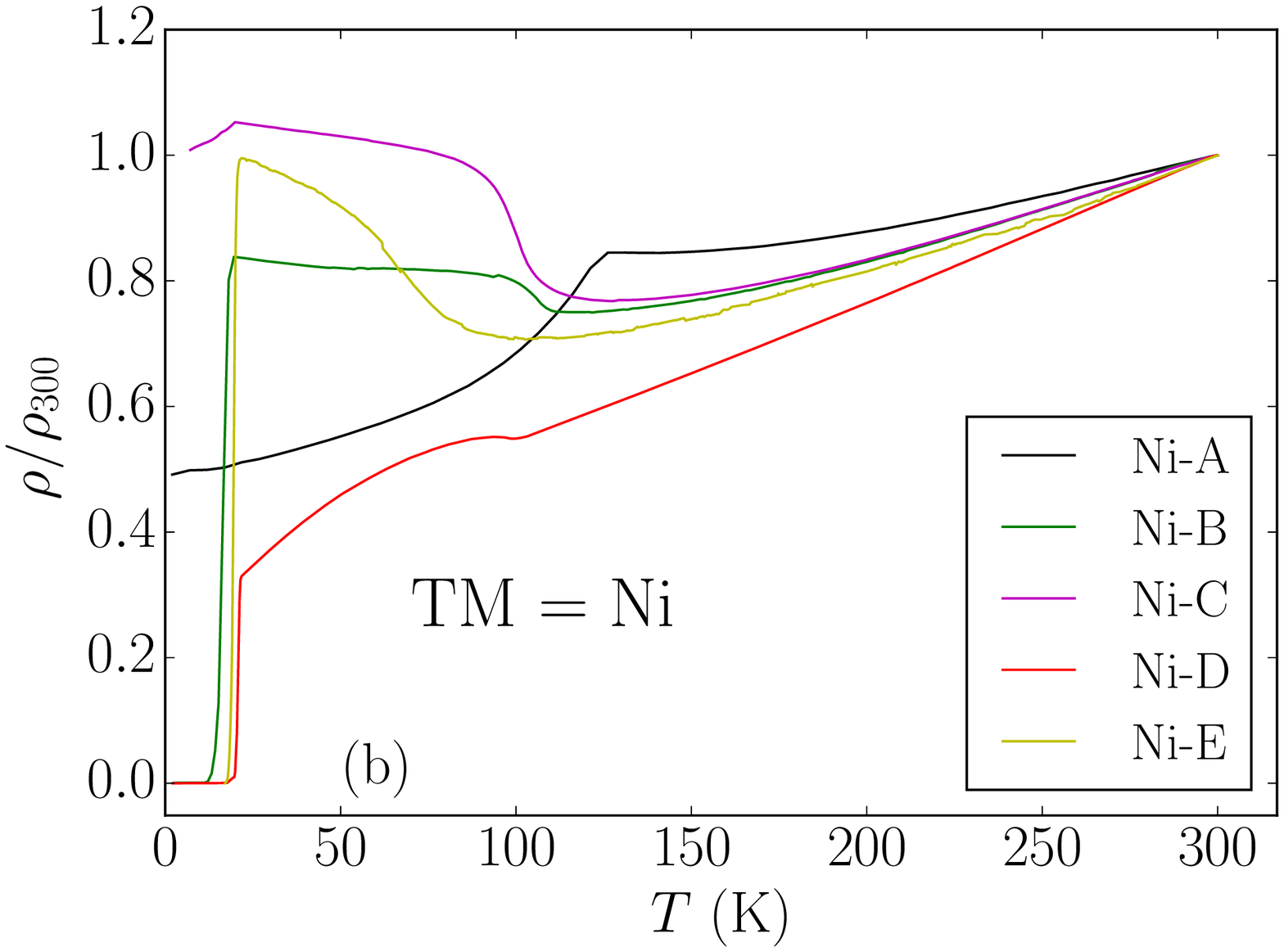}
\caption{Resistivity as a function of the temperature for the series of samples partially substituted with (a) TM = Co  and  (b) TM = Ni.
In both cases the resistivity is normalized by its value in $T=300$ K.} \label{RxT_CoNi}
\end{figure}

Panels (a) and (b) of Fig. \ref{RxT_CoNi} show the electrical resistivity as a function of the temperature for the series of 
samples with  TM = Co and TM = Ni, respectively.
A prominent feature in these results is the marked increase of the resistivity below the characteristic
temperature denoted as $T_{SDW}$. 
The observed hump reminds the superzone effect related to the opening of a gap in the conduction band due to the antiferromagnetic ordering
 ~\cite{A-mackintosh, B-elliott, C-miwa}. 
The gap and the associated Fermi surface distortion are expected to occur
when the periodicity of the magnetic lattice does not coincide with that of the atomic lattice.
This effect is commonly observed closely below the N\'eel temperature of antiferromagnetic metals, as Cr ~\cite{D-arajs}, Mn ~\cite{E-meaden} and some of the
heavy rare-earths ~\cite{F-legvold}. 
In the FeSC, that incommensurability probably occurs because of the nematicity observed together with the $SDW$ ordering ~\cite{paglione}. 
At this point it is worth to note that the superzone effect is not so evident in the BaFe$_2$As$_2$  parent compound.
In the BaFe$_2$As$_2$ the opening of a gap driving an effective reduction of the electronic density in the ordered phase tends to be compensated
by a higher mobility due to coherent scattering; thus, a non-metal behavior of the $\rho$ vs. $T$ curve is observed in a very small region.
In the substituted specimens, where the structural transition occurs at $T=T_S$ such that $T_S > T_{SDW}$ ~\cite{chu2009},
an increasing resistivity is already observed below $T_S$, and it continues to grow below $T_{SDW}$ until $T=T_c$. 
This implies that the TM substitution disrupts the commensurability of the $SDW$ ordering leading to an enhanced region where the 
superzone effect is observed in the substituted samples with respect to that of the pure compound.
In this scenario, the main role of the structural transition is to remove the commensurability of the magnetic and structural lattices
and this, together with the scattering by magnetic excitations, gives origin to the hump in the resistivity.

In the results of Fig. \ref{RxT_CoNi}, one observes that the Ni atoms depress $T_{SDW}$ more efficiently than Co in similar quantities. 
The qualitative comparison between results in Figs. \ref{RxT_CoNi}(a) and \ref{RxT_CoNi}(b) also suggests that Ni disturbs the coherent carrier 
scattering by magnetic excitations more than Co. Indeed, the resistivity in the BaFe$_{2-x}$Ni$_x$As$_2$ compounds show
a marked tendency to stay constant below $T_{SDW}$, whereas in the systems with \mbox{TM = Co} $\rho$ decreases neatly below the magnetic ordering 
temperature because of the continuous reduction of thermal induced spin disorder.
It should be noted, however, that  neither Co nor Ni introduce localized magnetic moments in BaFe$_{2-x}$TM$_x$As$_2$ \cite{rosa, garitezi2}. 
Then, the main role of both scatterers seems to  disrupt the $SDW$ ordered state characteristic of the parent 
compound.

\subsection{Magnetoresistance}

\begin{figure}
 \centering
\includegraphics[keepaspectratio,width =7.4truecm]{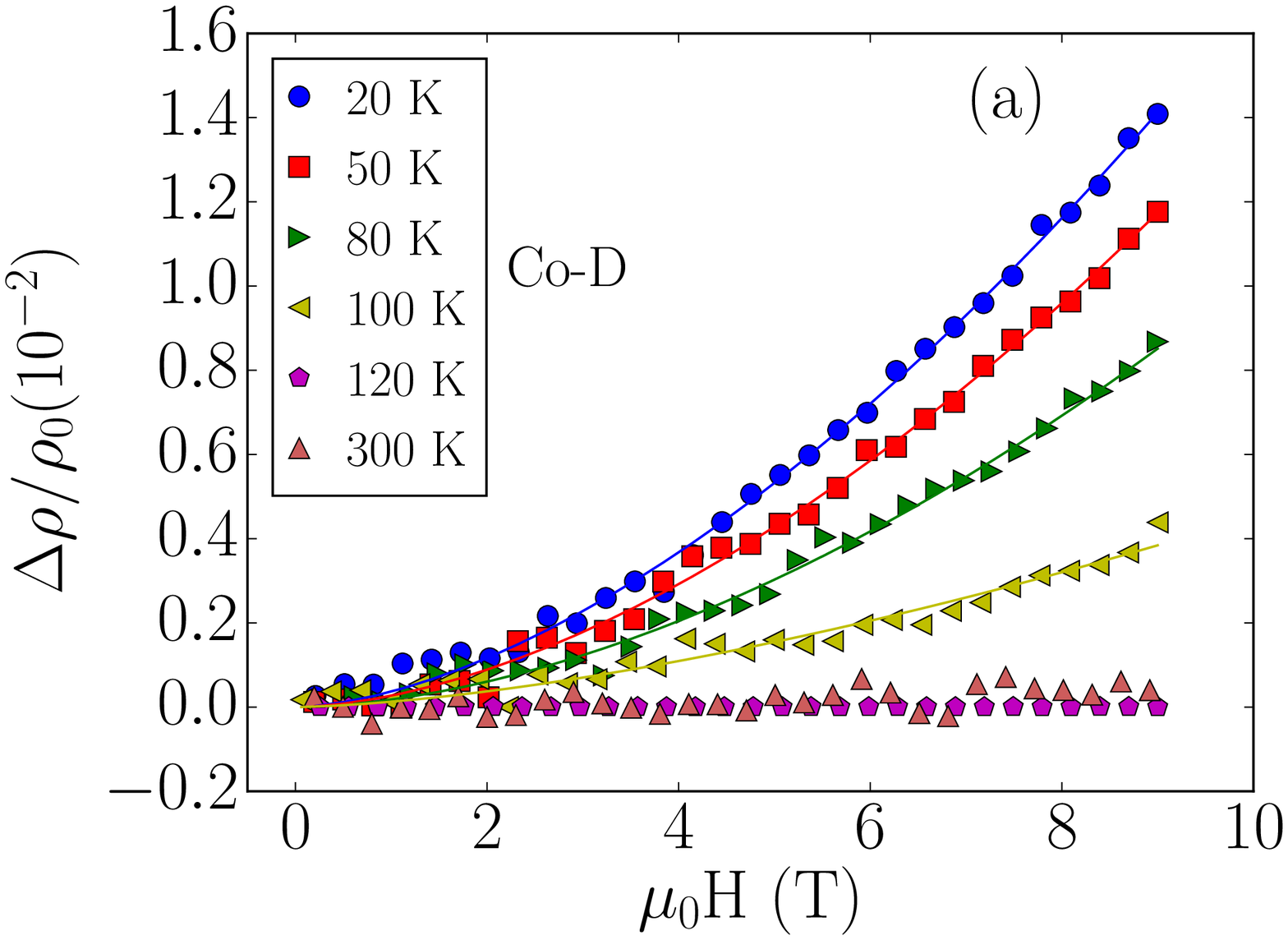}
\includegraphics[keepaspectratio,width =7.4truecm]{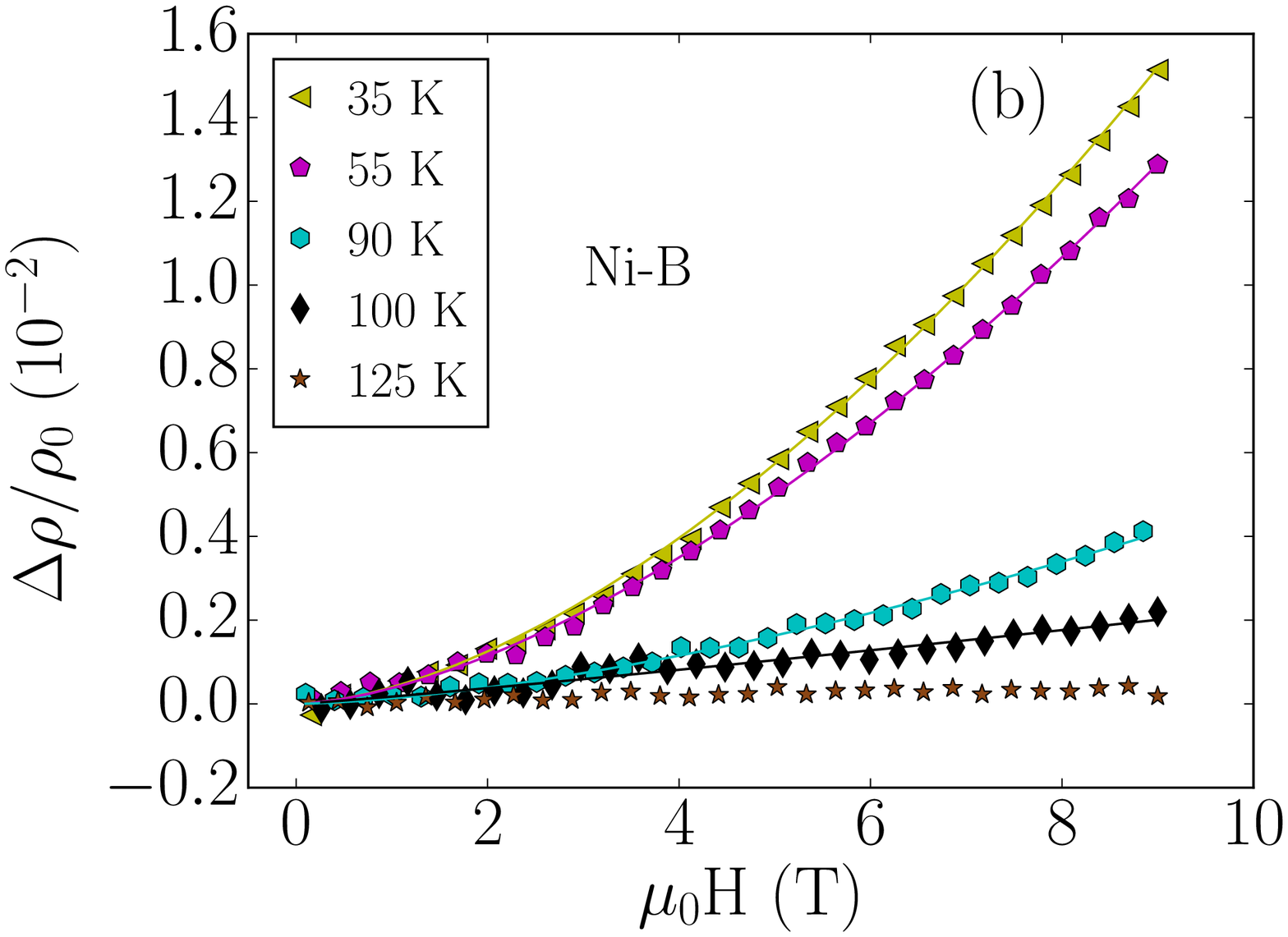}
\caption{Magnetoresistance as a function of the magnetic field in several fixed temperatures for samples Co-D ($x=0.043$) and Ni-B ($x=0.030$). 
Solid lines are fits to equation  $\Delta\rho=aB^b$, with $b \simeq 3/2$} \label{MR_B}
\end{figure}

The  magnetoresistance (MR) is given as the quotient \mbox{$\frac{\Delta\rho}{\rho(0)}=\frac{\rho(B)-\rho(0)}{\rho(0)}$.}
In Figs. \ref{MR_B}(a) and \ref{MR_B}(b), measurements of the MR as a function of  the magnetic-field 
in several fixed temperatures are shown for representative samples of the Co and Ni substituted series, respectively. 
The MR behaves as a power law of the applied field,  $\Delta\rho=aB^b$, with $b \simeq 3/2$, much like previously observed in slightly
substituted samples ~\cite{pena}. 

\begin{figure}
 \centering
\includegraphics[keepaspectratio,width =7.2truecm]{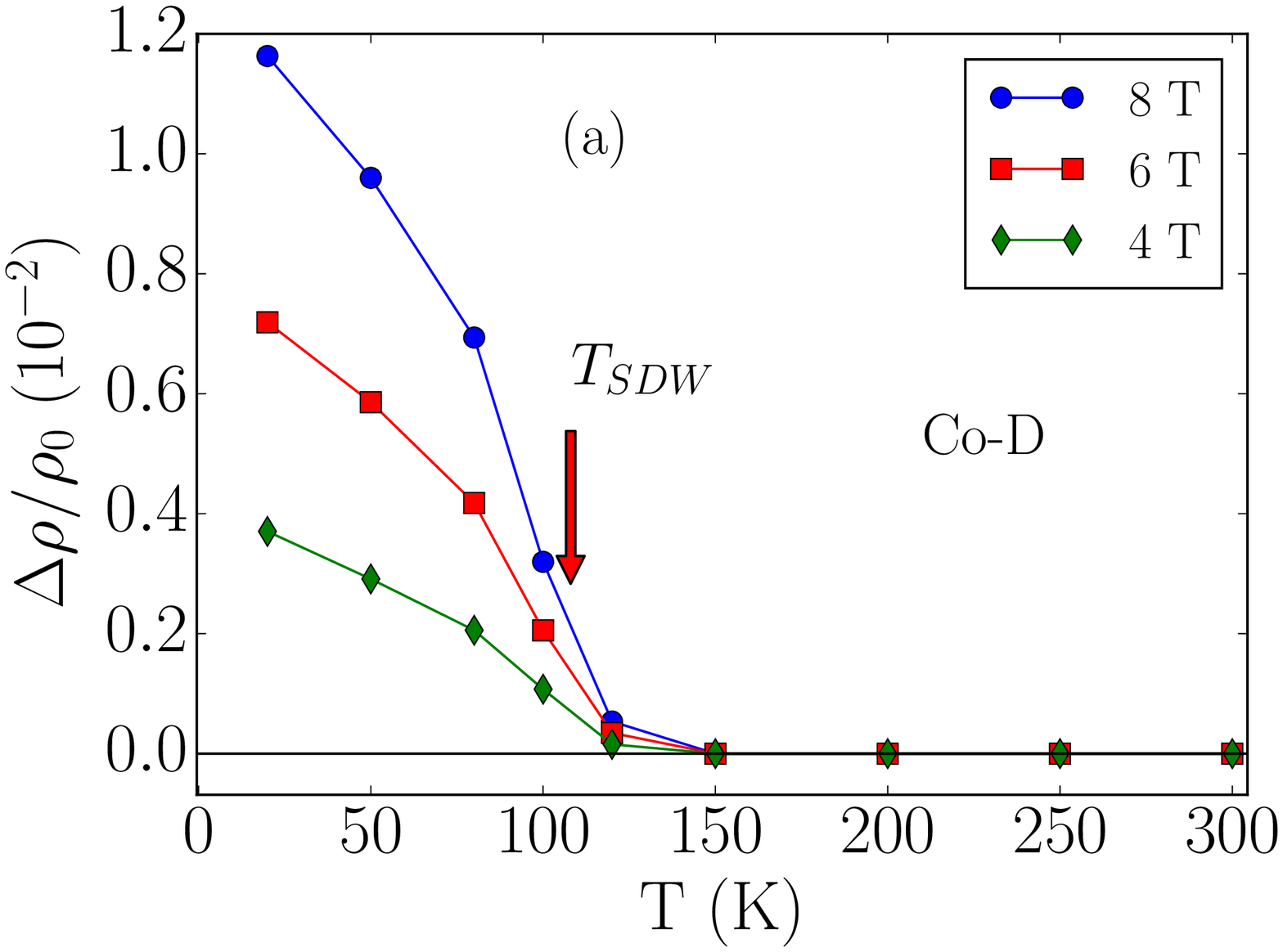}
\includegraphics[keepaspectratio,width =7.2truecm]{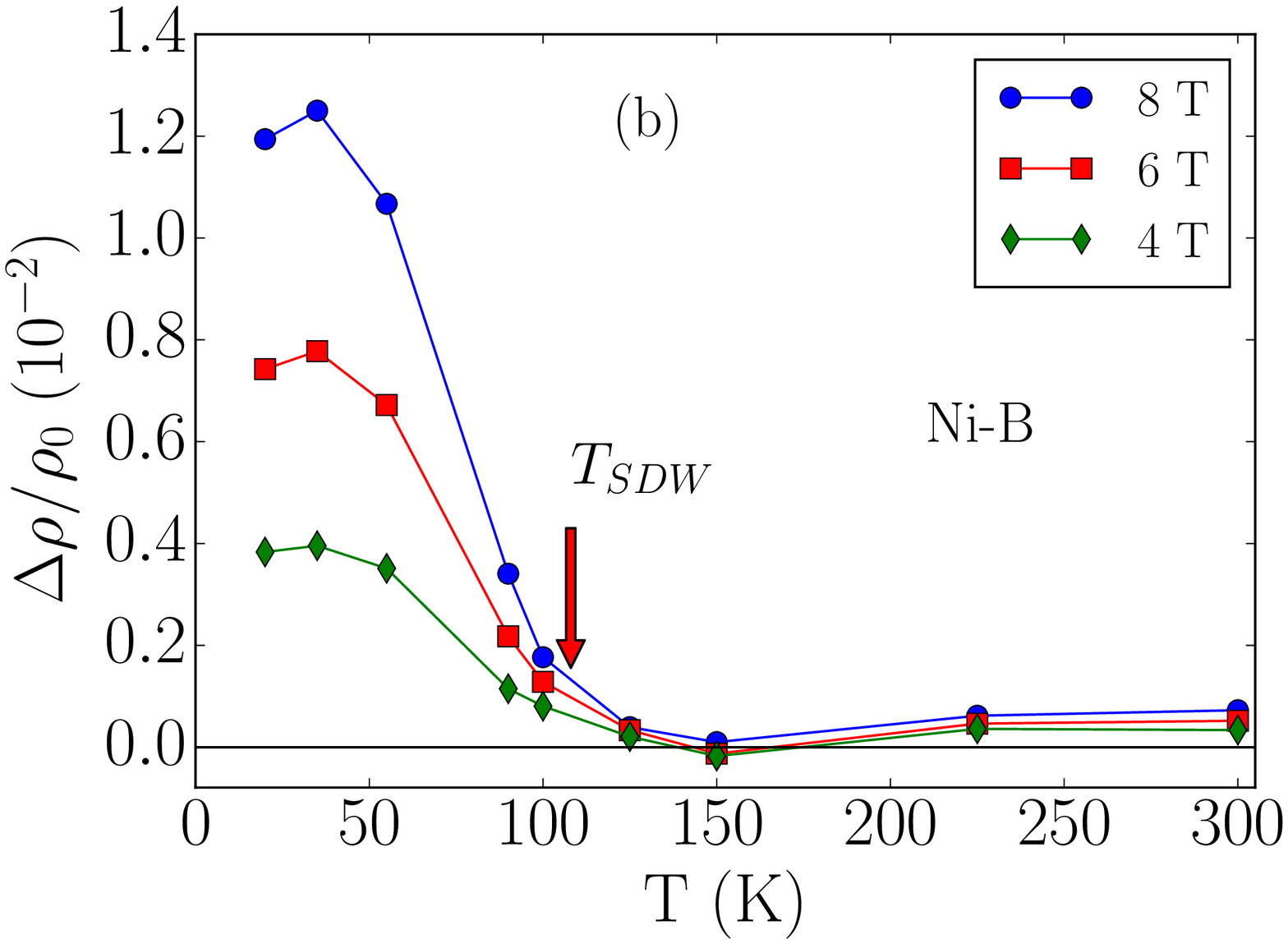}
\caption{MR amplitudes in several fixed fields plotted as a function of the temperature for samples Co-D ($x=0.043$) in panel (a)
and Ni-B ($x=0.030$) in panel (b)} \label{MR_T}
\end{figure}

In Figs. \ref{MR_T}(a) and \ref{MR_T}(b) the MR amplitude in three fixed fields is shown 
as a function of the temperature for the same samples of Fig. \ref{MR_B}. In both cases, one observes the striking resemblance of 
the MR amplitude with an order parameter which becomes measurable only below the transition temperature $T_{SDW}$.  

In the two-band model the MR is given by  ~\cite{ziman}:
\begin{equation}\label{eqMR}
 \Delta\rho\approx\frac{\sigma_h\sigma_e(\mu_h-\mu_e)^2}{(\sigma_h+\sigma_e)^2}\text{H}^2,
\end{equation}
 where $\sigma_{h(e)}$ is the  conductivity of the hole (electron) band and $\mu_{h(e)}$ is the respective mobility. 
The small MR observed above $T_{SDW}$ in both panels of Fig. \ref{MR_T} can be explained within this model assuming the presence of two types of
carriers with similar mobilities. This is a reasonable assumption taking into account experimental results
showing that both, electron and hole bands contribute equally  to the charge  transport in most of the  phase diagram
of the FeSC of the 122 family \cite{fang, olariu}.

On the other hand, the validity of Eq. (\ref{eqMR}) in the magnetically ordered state would require a MR quadratically dependent
on the field and a large difference between the mobilities of the hole and electron bands. 
The first requirement is not fulfilled by the experimental data.
The second would imply that abrupt and drastic changes have to occur in the Fermi surface of electron doped FeSC, 
with suppression of the hole pocket,  because of the magnetic ordering.
Indeed, a reconstruction of the FS occurs at $T = T_{SDW}$  because of the $SDW$ gap opening. 
However, this effect is not expected to produce strong changes in the hole and and electron mobilities ~\cite{shimojima2010}. 
On the other hand, it has been argued that a significant modification of the Fermi surface in electron doped FeSC,
with suppression of the hole pocket may occur because of a topological Lifshitz transition accompanying the magnetic
ordering at $T_{SDW}$. 
Again, this argument must be confronted to the fact that optical spectroscopy and ARPES measurements suggest that the proposed
Lifshitz transition occur only in pure and heavily underdoped samples of the 122 Fe-pnictides system ~\cite{marsik2013, dhaka2011, liu2010}. 
Thus, assuming that holes are removed because of a Lifshitz transition and MR is enhanced according to Eq. (\ref{eqMR}) would
leave the behavior of the magneto-resistance unexplained in most of our samples, which are not in the TM concentration
limit where the topological effect have been proposed to set in (this is true independently of TM being Co or Ni). 
Moreover, since there's no order parameter associated with a topological Lifshitz transition, to explain the temperature 
dependence of the MR as shown in Fig. \ref{MR_T} solely with basis on this concept does not follow straightforwardly. 
Finally, even considering that the main changes in the electronic structure observed in spectroscopic measurements
can be taken as originated by interband phenomena developing near the $\Gamma-$pocket, one can not not exclude the 
possibility that the charge carriers are strongly interacting with magnetic, orbital, or nematic fluctuations ~\cite{marsik2013}.
These facts suggest that an additional and quantitatively more relevant mechanism, other than the multiple-band conduction,
must be taken into consideration to fully explain the magneto-transport properties of the 122 Fe-pnictides. 
We then propose that electron scattering by magnetic excitations have an important role to explain the MR in 
the magnetically ordered region of our samples.

\begin{figure}
 \centering
\includegraphics[keepaspectratio,width =8.2truecm]{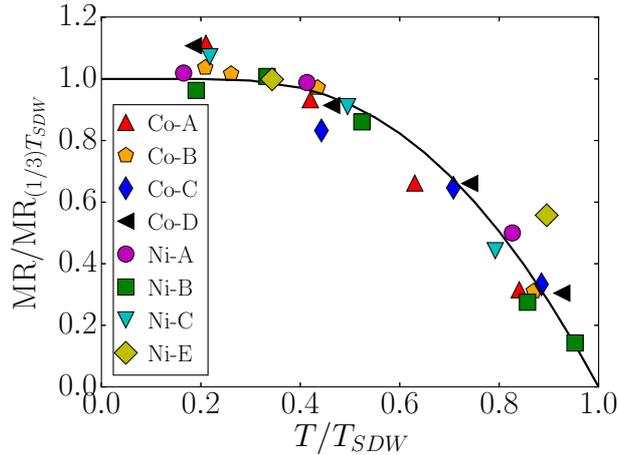}
\caption{MR amplitude normalized to its value at $T=(1/3)T_{SDW}$  as a function of the normalized temperature
$T / T_{SDW}$ for the Co and Ni-substituted series of samples. The solid line represents the square of the Brillouin function
for $J=1/2$ (see text).} \label{MR_nor20}
\end{figure}

In order to seek for experimental evidences supporting our assumption, in Fig. \ref{MR_nor20}
we plot the MR amplitude at $B = 8$ T,  normalized to its value at $T =\frac{1}{3}T_{SDW} $,  as a function of the reduced temperature 
$T / T_{SDW}$  for the both series of samples studied here. 
The black, continuous line represents the square of the normalized staggered magnetization as a function of the temperature,
as described by the Brillouin function for $J=1/2$. 
This value for $J$ coincides with the value calculated  for the effective 
local spin,  $S_{eff}$,  in the spin-state crossover model in Ref. ~\cite{chaloupka}.
That effective spin arises from the dynamical mixing of the quasidegenerate spin states of Fe$^{2+}$ ions 
by intersite electron hoppings  ~\cite{chaloupka}. 
The collapse of the data for both families of samples over the line 
predicted by the mean field theory is remarkable, and 
suggest that this theory is roughly valid and the magnetorresistance is proportional
to the square of the staggered magnetization. 
Moreover, this clearly indicates a universal scaling between the MR  value with $T_{SDW}$ in the FeSC of the 122 family.
There are however two exceptions to this universal scaling, these are the samples Co-E e Ni-D. 
We argue this can indicate that the effects of the magnetic ordering in the resistivity and MR are quite weak for these two specimens.

\subsection{Hall effect}

\begin{figure}
 \centering
\includegraphics[keepaspectratio,width =7.2truecm]{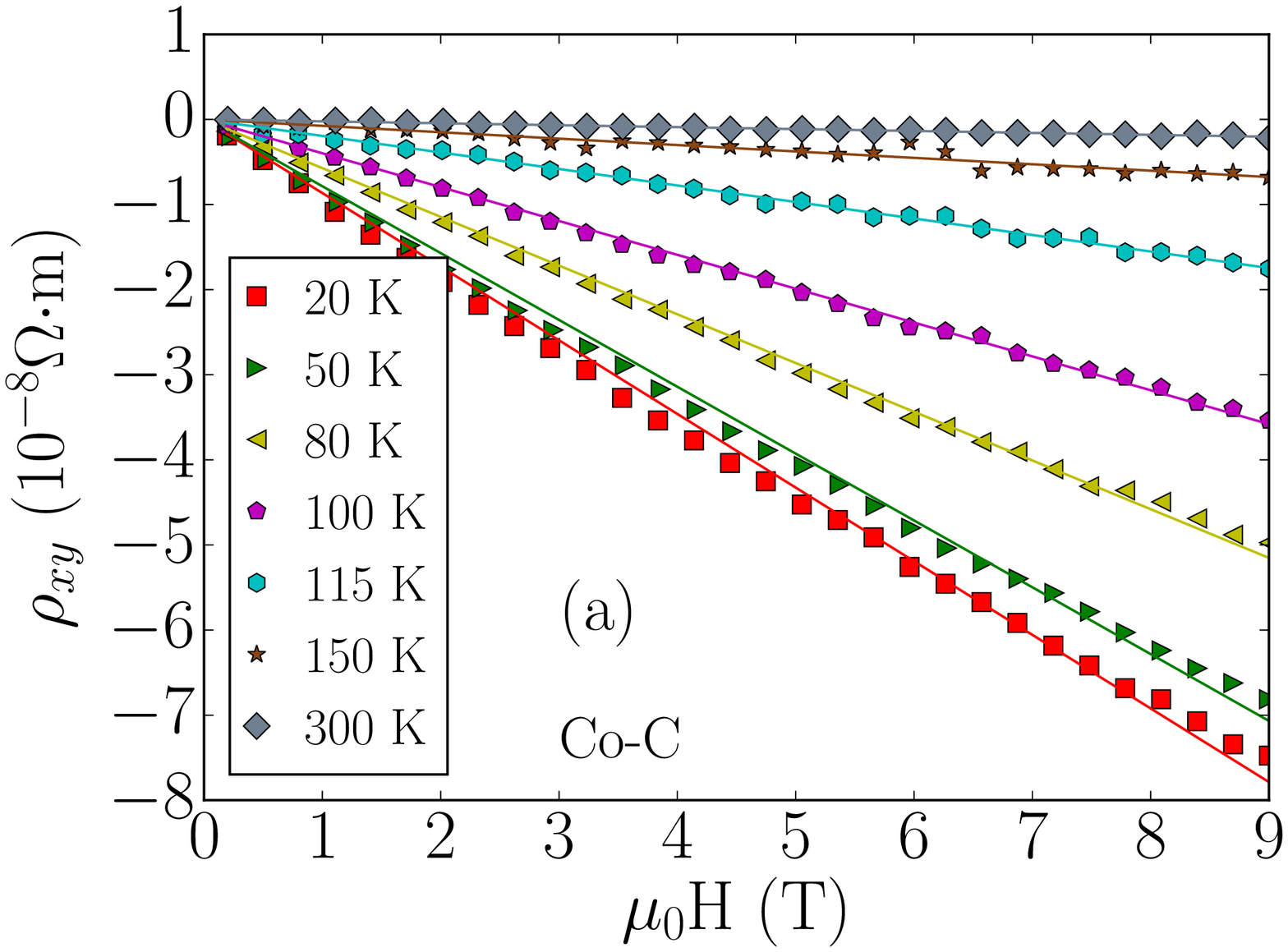}
\includegraphics[keepaspectratio,width =7.2truecm]{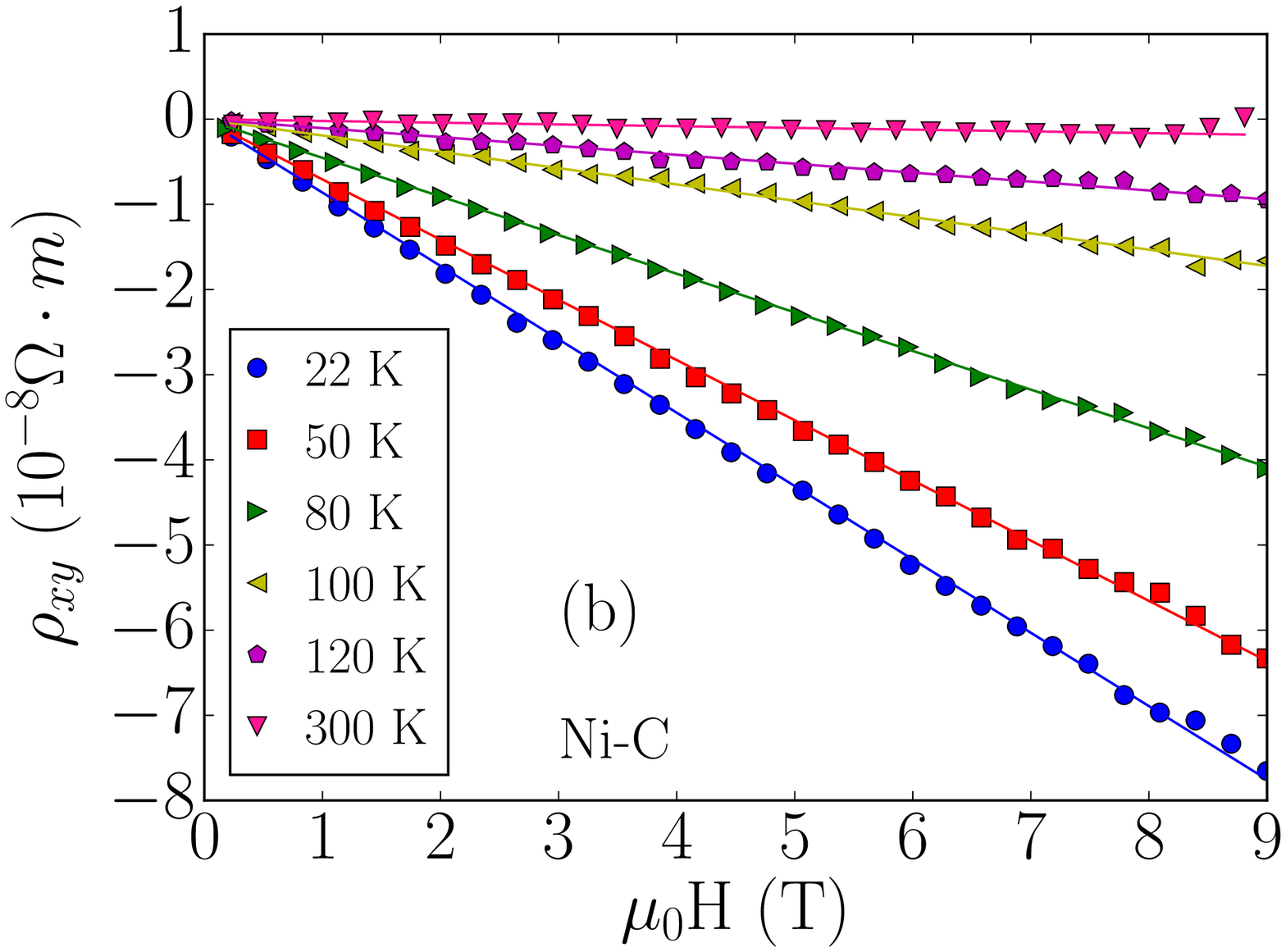}
\caption{Hall resistivity as a function of the applied field for the samples Co-C ($x=0.037$) and Ni-C ($x=0.033$), 
shown in panels (a) and (b), respectively. The experimental data are fitted to straight lines} \label{rxy_vs_H}
\end{figure}

In Fig. \ref{rxy_vs_H} we show representative transversal resistivity curves ($\rho_{xy}$ vs. $H$) for two of the studied samples
of BaFe$_{2-x}$TM$_x$As$_2$. Namely, results for samples Co-C ($x=0.037$) and Ni-C ($x=0.033$) are shown in panels (a) and (b), respectively. 
The same linear behavior of $\rho_{xy}$ as a function of the field was observed for all samples in the complete temperature 
interval investigated (except for measurements performed very close the superconducting critical temperature). 
Thus, in all cases the Hall coefficient  $R_H$ was calculated as the slope of those straight lines.

\begin{figure}
 \centering
\includegraphics[keepaspectratio,width =7.2truecm]{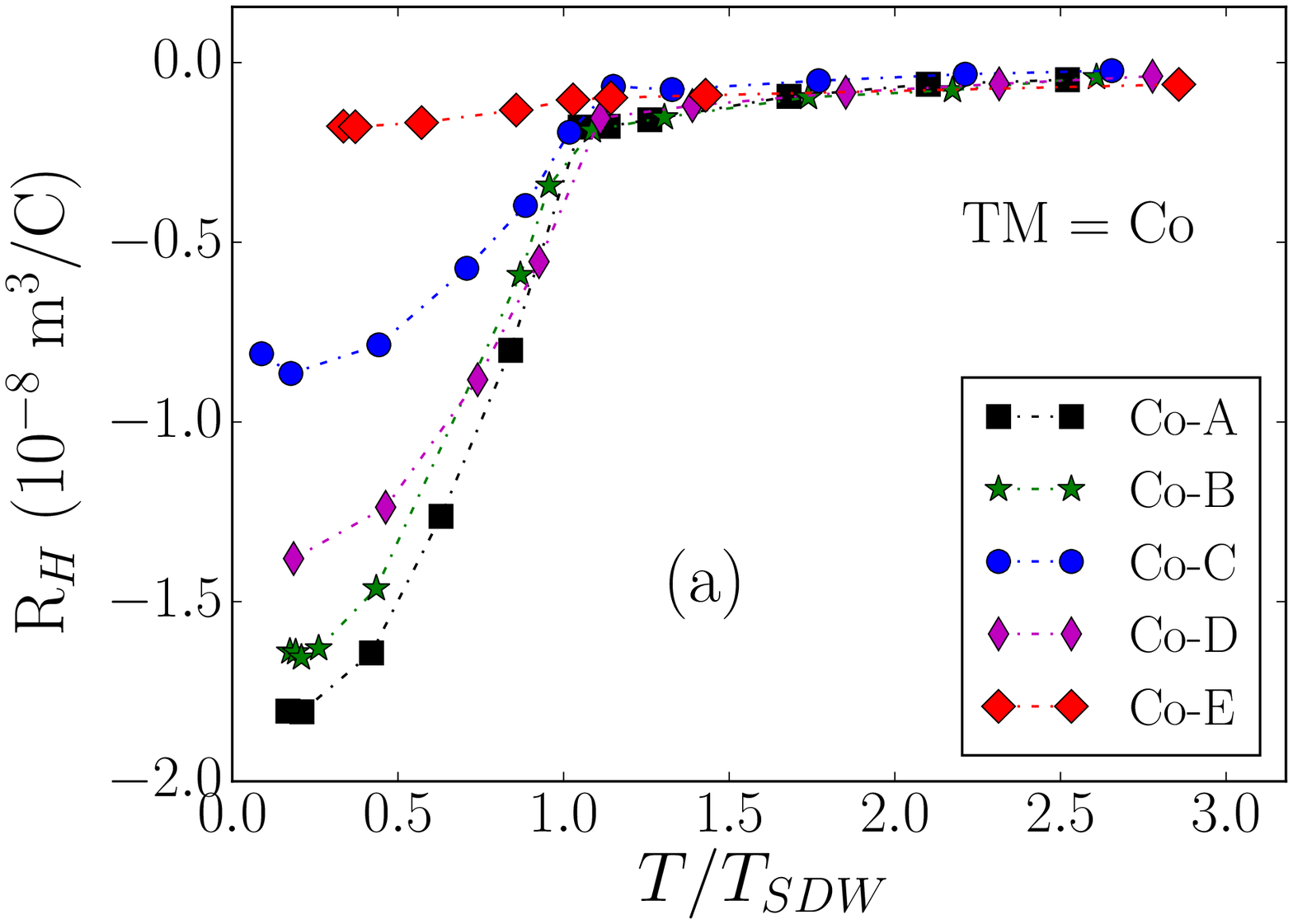}
 \includegraphics[keepaspectratio,width =7.2truecm]{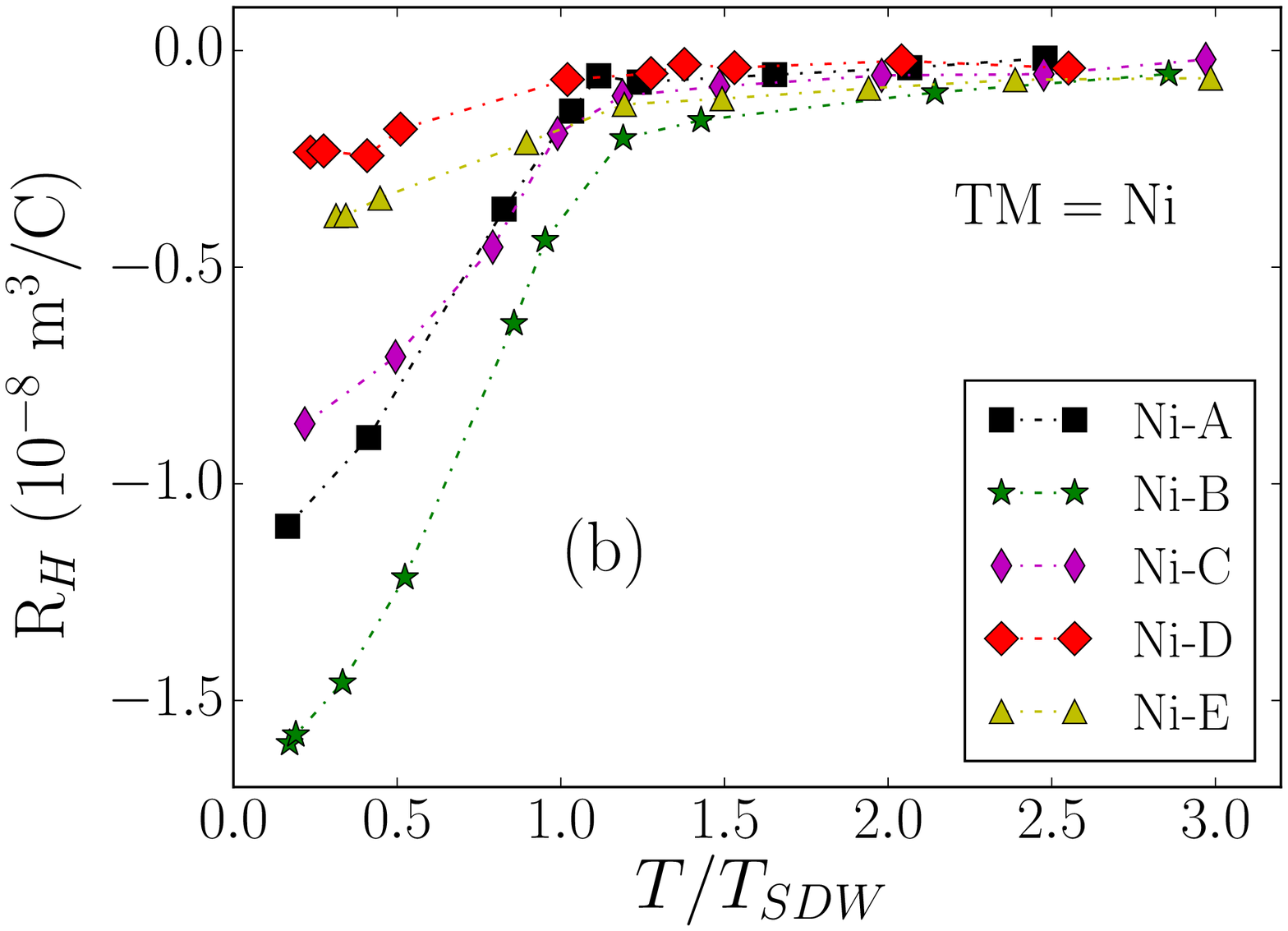}\\
\caption{Hall coefficient as a function of the temperature for samples with TM = Co and TM = Ni in panels (a) and (b) respectively. 
The values for $R_H$ were extracted from the slope of linear fittings of plots as those shown in Fig. \ref{rxy_vs_H}.} \label{RH_vs_T}
\end{figure}

The Hall coefficient $R_H$ is presented as a function of the temperature in Fig. \ref{RH_vs_T} for 
the series of samples where TM = Co in panel (a) and  TM = Ni in panel (b). 
The most conspicuous characteristic of curves in Fig. \ref{RH_vs_T} is the strong change in the magnitude and temperature 
dependence of $R_H$ at $T = T_{SDW}$. Above $T_{SDW}$ the Hall coefficient is small and weakly temperature dependent. 
Below $T_{SDW}$, the magnitude of $R_H$  increases sharply. 
As for the MR, the small value of $R_H$ in the PM regime can be explained by a simple two-band model
with two almost compensated bands of similar mobilities, as expressed by the equation ~\cite{ziman}: 
\begin{equation}\label{eqRH_2b}
 R_H\approx\frac{\sigma_h\mu_h-\sigma_e\mu_e}{(\sigma_h+\sigma_e)^2}.
\end{equation}
Once again, assuming the hypothesis that the two-band model is enough to explain the strongly different magnitude and temperature
dependence of $R_H$ in the magnetically ordered region demands a drastic reduction in the hole-type carrier density, 
as assumed in Ref. ~\cite{albenque_lett}, or significant and opposite changes in the mobilities of electrons and holes,
as supposed in Refs. ~\cite{fang} and ~\cite{olariu}. 
Indeed, it might be tempting to consider that a severe Fermi surface reconstruction (Lifshitz transition)
occurring at $T = T_{SDW}$ triggers a large and continuous change in the magnitude of the carriers density in the magnetically ordered phase
of the Fe-pnictides when compared to that of the PM state. For example, ARPES measurements and theoretical calculations
in Ref. ~\cite{brouet2013} support the existence of an unexpected temperature dependence of the Fermi surface geometry 
in BaFe$_{2-x}$Co$_x$As$_2$ samples. 
However, the study in Ref. ~\cite{brouet2013} is focused on samples corresponding to the overdoped side of the dome.
These samples do not present a magnetically ordered phase. Moreover, results reported by these authors for 
the undoped compound do not hint on evidences for a severe lost of carriers at $T = T_{SDW}$.
Conversely, the number of carriers obtained by them at $T \sim 0$ and $T = 200$ K is almost the same ~\cite{brouet2013}.

Other experimental studies, which analyze samples covering a larger region of the phase diagram and focus
on the differences among the properties of the PM and $SDW$ phases, conclude that a Lifshitz transition occurs in 
specimens where the TM content is about or below that characterizing the emergence of the superconducting dome ~\cite{liu2010, deok2009}. 
However, ours and other 
Hall effect results  ~\cite{albenque_lett, fang, olariu} show that the qualitative behavior of the $R_H$ vs. $T$ curves is preserved whenever 
a magnetically ordered phase is observed, independently of the TM content. Therefore, the hypothesis evoking the 
lost of carriers due to the occurrence of a Lifshitz transition at $T = T_{SDW}$ may not apply to explain the sharp increase 
in the $R_H$ magnitude observed below this point in samples spanning a large part of the phase diagram. 
As for the MR, the above considerations lead us to propose that the interaction of the charge carriers with magnetic 
excitations should also be considered to fully describe the Hall effect behavior in the FeSC where an intermediate 
magnetic phase is present. In other terms, the Hall coefficient should have its temperature dependence attributed 
to an anomalous term of magnetic origin that comes into play in temperatures below $T_{SDW}$.

\begin{figure}
 \centering
  \includegraphics[keepaspectratio,width =7.2truecm]{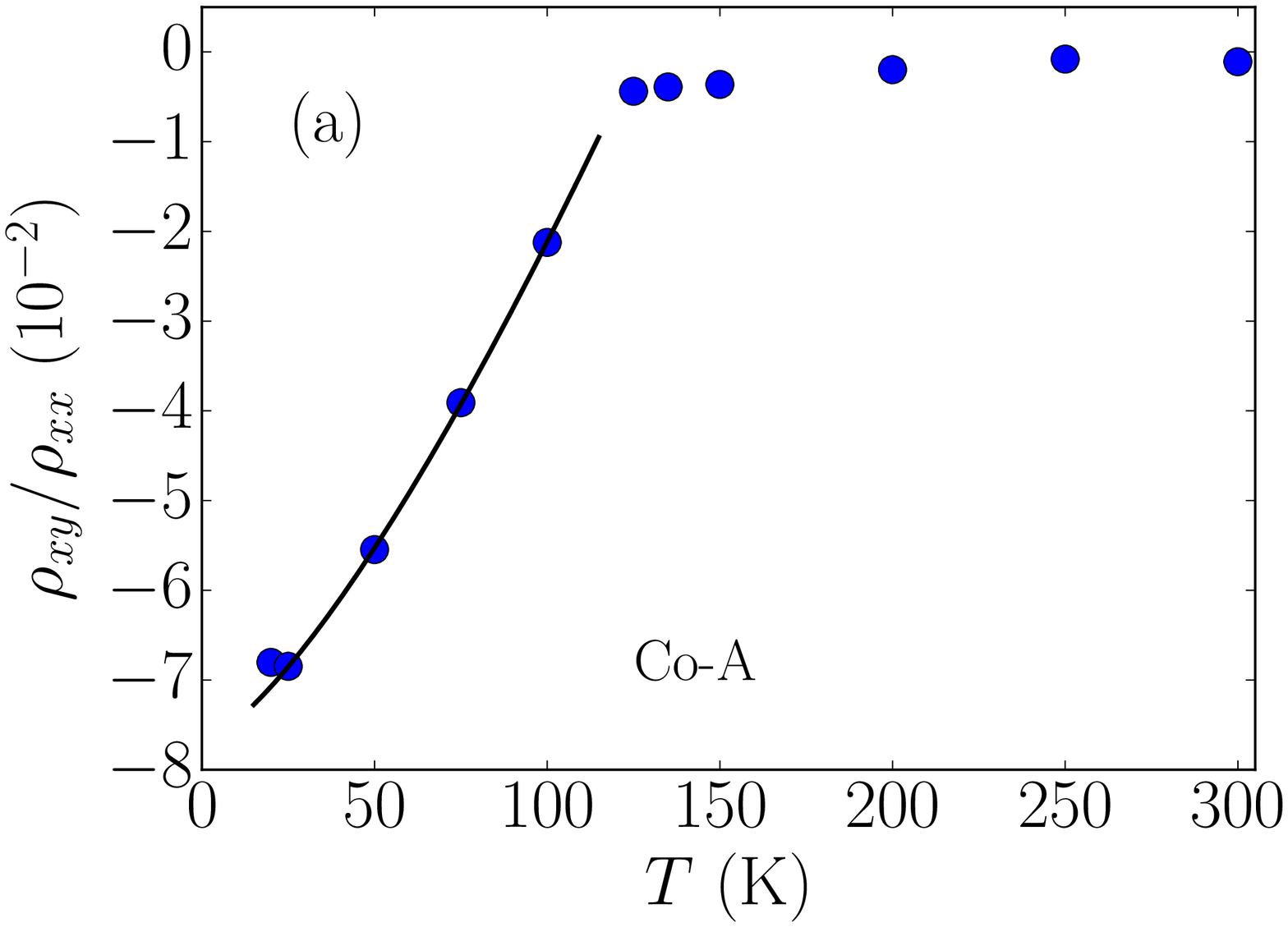}
     \includegraphics[keepaspectratio,width =7.2truecm]{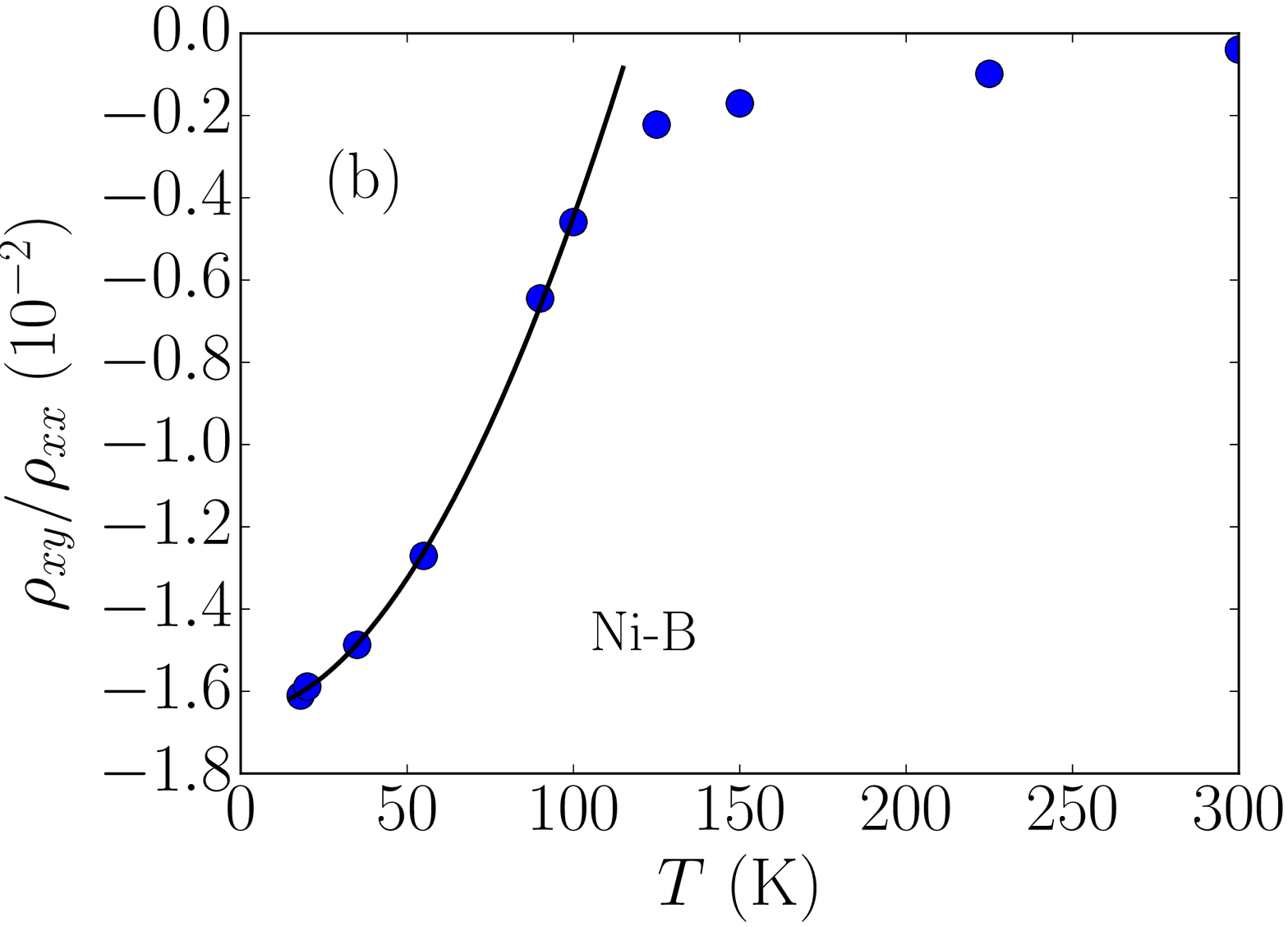}
 \caption{Tangent of the Hall angle for the samples (a) Co-A ($x=0.023$) and (b) Ni-B ($x=0.030$). Both data sets are  for a fixed
 magnetic field of magnitude 4 T.} \label{Tan_vs_T_CoNi}
\end{figure}

As a final insight, in panels (a) and (b) of Fig. \ref{Tan_vs_T_CoNi} we plot the tangent of the Hall angle,
($\tan\Theta_H=\rho_{xy}/\rho_{xx}$), as a function of the temperature for representative Co-substituted and Ni-substituted samples, respectively. 
Fittings of the Hall angle to a function of the form 
\begin{equation}\label{eqTanT}
   \tan\Theta_H=\alpha+\beta T^n
\end{equation}
are presented as solid black lines and the fitting coefficients and exponents are listed for all samples in \mbox{Table \ref{TanTheta_fit_coef}}.
Results in Fig. \ref{Tan_vs_T_CoNi} may be compared to those published in Ref. ~\cite{pena}. One then observes that $\tan\Theta_H(T)$ evolves 
from a linear function of the temperature in the slightly substituted, and non-superconducting, samples ~\cite{pena} 
to an approximately quadratic temperature behavior in samples with higher TM content, for which a superconducting
ground state is stabilized. It is also shown in the results of Fig.  \ref{Tan_vs_T_CoNi} that $\tan\Theta_H$ tends to zero as the superconducting 
transition is approached from above.  

From data in Table \ref{TanTheta_fit_coef} one observes: (i) when $T\rightarrow0$ there's a non-zero 
contribution to the Hall tangent represented by the parameter $\alpha$; 
(ii) the parameter $\alpha$ is, in general, higher in samples with lower TM content, (iii) the absolute values
for $\alpha$ are higher for the Co-substituted samples than for the Ni-substituted ones.
In Table \ref{TanTheta_fit_coef} it is also possible to see  that the
parameter $\beta$ almost vanishes in samples with the weakest magnetic order in both series of samples.
The first point strongly suggest that a non-collinear spin structure 
or some quenched magnetic disorder is present in the investigated samples. This contribution is normally associated
with spin chirality  ~\cite{nagaosa, sinitsyn}. The second and third points suggest that the chiral mechanism, present in samples 
with the most robust magnetic ordering, is diminished when that ordering is disturbed via the augmentation of the TM content.
The temperature dependent term in Eq. (\ref{eqTanT}) is rather related to skew scattering produced by individual spins  ~\cite{nagaosa, sinitsyn}. 
This term also shows a tendency to vanish when $x$ is increased, thus revealing the gradual weakening of the intermediate
magnetic order as the TM content increases. We thus propose that the temperature and $x$ dependence of $\tan\Theta_H$ in the
BaFe$_{2-x}$TM$_x$As$_2$ compounds are related to anomalous contribution to the Hall effect that have origin in non-collinear
and localized magnetic moments  that coexists with the itinerant ones ~\cite{zhao, liu, 117dai} responsible by the $SDW$ arrangement.

\begin{table}
\caption{Parameters obtained from fittings of $\tan\Theta_H$ to Eq. (\ref{eqTanT}) in the region $T < T_{SDW}$ for
the both series of studied samples}\label{TanTheta_fit_coef}
\small
\centering
\begin{tabular}{cccc|cccc}
 &  \textbf{TM = Co} & & & & \textbf{TM = Ni} & & \\
 \hline
 $x$ & $\alpha\times10^{-3}$ & $\beta\times10^{-5}$ & $n$ & $x$ & $\alpha\times10^{-3}$ & $\beta\times10^{-5}$ & $n$\\
  \hline
 0.023 	   & -77$\pm$1 & 10.00 $\pm$ 2.0 & 1.36$\pm$0.04 & 0.015 & -31$\pm$1  & 13.0 $\pm$   1.00 & 1.13$\pm$0.01 \\
 0.032     & -40$\pm$1 & 0.90 $\pm$ 0.4 & 1.80$\pm$0.10  & 0.030 & -16$\pm$1   &    0.12   $\pm$  0.09 & 1.90$\pm$0.10\\
 0.037     & -78$\pm$2 & 4.00 $\pm$ 2.0 & 1.60$\pm$0.10  &  0.033 & -11$\pm$1  &    0.08 $\pm$  0.01 & 1.95$\pm$0.01\\
 0.043     & -53$\pm$1 & 0.80 $\pm$ 0.5 & 1.80$\pm$0.10  & 0.035 & -15$\pm$1   &    0.27   $\pm$  0.04 & 2.00$\pm$0.10\\
 0.118     & -6$\pm$1 & 0.03 $\pm$ 0.0	& 2.10$\pm$0.01 & 0.051 & -5$\pm$1 & 0.05 $\pm$ 0.01 & 2.00$\pm$0.10\\

%
%
	   
\end{tabular}
\end{table}

\section{Final comments and conclusions}

The importance of the scattering by magnetic fluctuation to the transport properties of Fe-based 122 pnictides, 
superconducting or not, has been pointed out in previous works ~\cite{pena, fernandes2011,  wang2015}. For example, the anisotropic 
resistance in the nematic phase has been described with an Ising-nematic model where the anisotropy is the product 
of the interference between scattering by impurities and by critical spin fluctuations ~\cite{fernandes2011}. Such a model emerges 
from the combination of magnetic fluctuations and frustration  ~\cite{fernandes2011}. 
On the other hand, multiple-bands conduction models
reproduce experimental resistance curves of FeSC in the absence of an external magnetic field ~\cite{eom, ishida}. 
Nevertheless these models are not able to adequately describe the magnetotransport properties in the $SDW$ 
phase of those compounds without assuming restrictive hypothesis on the mobility and/or number of the charge carriers. 
Thus, it seems that considering that the multiple-band models are able to fully describe the magnetotransport
properties of the FeSC is not entirely justified on experimental grounds. 
Lifshitz transitions and exotic temperature-dependent Fermi surfaces 
explain only partially the behavior of the MR and Hall coefficient. 
Our previous results in slightly doped and non-superconducting samples ~\cite{pena} as well as  the results of this work in samples
having a superconducting ground state while preserving an intermediate magnetic phase, reveal that once a magnetic
transition occurs, large changes in the absolute values and temperature dependence of MR and the Hall coefficient 
are observed in the region $T \leq T_{SDW}$, independently of the TM substitution, or how low $T_{SDW}$ is.

Accumulative experimental evidences show that Fermi surface reconstructions are only observed in the pure and 
slightly substituted samples ($\sim20$ \%) ~\cite{dhaka2011, dhaka2013}.
On the other hand, altough temperature dependent energy bands have been considered as an intrinsic characteristic of the 122 
system, this peculiarity was not related with the existence of a magnetically ordered phase ~\cite{dhaka2013}. 
Then, this effect may not be considered as the unique mechanism to explain the striking differences of 
the magnetotransport properties of underdoped samples above and below $T_{SDW}$.
Taking into account that the iron-pnictogen distance controls the overlap between iron and pnictogen orbitals, enforcing iron electrons 
to become more localized with increasing distance  ~\cite{yin2011}, the fact that the $c-$axis lattice parameter decrease 
with addition of TM atoms in the doped compounds implies that itinerancy increases with doping. 
In the context of scattering by magnetic fluctuations, this scenario agrees with the fact that in pure and slightly doped samples,
a much stronger variation of MR and $R_H$ is observed below $T_{SDW}$ than in optimally doped and overdoped samples. 
This happens because of the more localized character of magnetic moments in small $x$ limit. 
We can then speculate  that the itinerant electrons are rather responsible for the multiple-band effects in the electron 
transport properties of the 122-FeSC compounds while more localized electrons are responsible for the anomalous MR and Hall effect components.

As a summary,  we have studied two series of BaFe$_{2-x}$TM$_x$As$_2$ crystals, 
where TM = Co and TM = Ni.
While presenting an intermediate magnetic state, most of our samples have a superconducting ground state. 
The superconducting transition temperatures do not always match with the magnetic ordering temperature $T_{SDW}$ for the dominant phase, 
indicating that some phase separation occurs in the studied samples.

We interpret the characteristic hump of the resistivity occurring at $T_{SDW}$  as a superzone effect,
which means that not only some reconstruction of the Fermi surface occurs at this temperature due to the stabilization 
of an antiferromagnetic-like state, but intense scattering by magnetic excitations plays a major role. 
We also proposed that the sharp increase of the MR amplitude observed in the magnetically ordered phase, 
and its universal scaling with the reduced temperature $T / T_{SDW}$, indicates that this magneto-transport property 
is rather related with scattering of carriers by magnetic excitations than with drastic  changes in the mobility or density 
of the charge carriers. Finally, our Hall effect results suggest that anomalous terms related to magnetic excitations 
contribute to the Hall resistivity of all the studied samples in the magnetically ordered phase. 
This contribution is progressively weakened upon increasing the Co or Ni concentration.

\section*{Acknowledgments}

This work was partially financed by the Brazilian agencies FAPERGS (Grant PRONEX 12/2014),
FAPESP (Grants  2012/05903-6, 2012/04870-7 and \mbox{2011/01564-0}),
and CNPq (Grants PRONEX 12/2014, 304649/2013-9 and 442230/2014-1).
J. P. Pe\~na benefits from a PNPD/CAPES fellowship.

\end{doublespace}


\begin{thebibliography}{}%
\makeatletter
\providecommand \@ifxundefined [1]{%
 \ifx #1\undefined \expandafter \@firstoftwo
 \else \expandafter \@secondoftwo
\fi
}%
\providecommand \@ifnum [1]{%
 \ifnum #1\expandafter \@firstoftwo
 \else \expandafter \@secondoftwo
\fi
}%
\providecommand \enquote [1]{``#1''}%
\providecommand \bibnamefont  [1]{#1}%
\providecommand \bibfnamefont [1]{#1}%
\providecommand \citenamefont [1]{#1}%
\providecommand\href[0]{\@sanitize\@href}%
\providecommand\@href[1]{\endgroup\@@startlink{#1}\endgroup\@@href}%
\providecommand\@@href[1]{#1\@@endlink}%
\providecommand \@sanitize [0]{\begingroup\catcode`\&12\catcode`\#12\relax}%
\@ifxundefined \pdfoutput {\@firstoftwo}{%
 \@ifnum{\z@=\pdfoutput}{\@firstoftwo}{\@secondoftwo}%
}{%
 \providecommand\@@startlink[1]{\leavevmode\special{html:<a href="#1">}}%
 \providecommand\@@endlink[0]{\special{html:</a>}}%
}{%
 \providecommand\@@startlink[1]{%
  \leavevmode
  \pdfstartlink
   attr{/Border[0 0 1 ]/H/I/C[0 1 1]}%
   user{/Subtype/Link/A<</Type/Action/S/URI/URI(#1)>>}%
  \relax
 }%
 \providecommand\@@endlink[0]{\pdfendlink}%
}%
\providecommand \url  [0]{\begingroup\@sanitize \@url }%
\providecommand \@url [1]{\endgroup\@href {#1}{\urlprefix}}%
\providecommand \urlprefix [0]{URL }%
\providecommand \Eprint[0]{\href }%
\@ifxundefined \urlstyle {%
  \providecommand \doi [1]{doi:\discretionary{}{}{}#1}%
}{%
  \providecommand \doi [0]{doi:\discretionary{}{}{}\begingroup
  \urlstyle{rm}\Url }%
}%
\providecommand \doibase [0]{http://dx.doi.org/}%
\providecommand \Doi[1]{\href{\doibase#1}}%
\providecommand \bibAnnote [3]{%
  \BibitemShut{#1}%
  \begin{quotation}\noindent
    \textsc{Key:}\ #2\\\textsc{Annotation:}\ #3%
  \end{quotation}%
}%
\providecommand \bibAnnoteFile [2]{%
  \IfFileExists{#2}{\bibAnnote {#1} {#2} {\input{#2}}}{}%
}%
\providecommand \typeout [0]{\immediate \write \m@ne }%
\providecommand \selectlanguage [0]{\@gobble}%
\providecommand \bibinfo [0]{\@secondoftwo}%
\providecommand \bibfield [0]{\@secondoftwo}%
\providecommand \translation [1]{[#1]}%
\providecommand \BibitemOpen[0]{}%
\providecommand \bibitemStop [0]{}%
\providecommand \bibitemNoStop [0]{.\EOS\space}%
\providecommand \EOS [0]{\spacefactor3000\relax}%
\providecommand \BibitemShut [1]{\csname bibitem#1\endcsname}%
\end{thebibliography}%


\begin{thebibliography}{}
\bibitem{paglione} J. Paglione and R. Greene, Nature Physics 6 (2010) 645-658
\bibitem{kishida} K. Ishida, Y. Nakai, and H. Hosono, JPSJ 78 (2009) 062001
\bibitem{hsueh} Hsueh-Hui Kuo \textsl{et. al.}, Phys. Rev. B 84 (2011) 054540
\bibitem{kkhuynh} K. K. Huynh, Y. Tanabe, and K. Tanigaki, Phys. Rev. Lett. 106 (2011) 217004
 \bibitem{albenque_lett} F. R. Albenque, D. Colson, A. Forget and H. Alloul, Phys. Rev. Lett. 103 (2009) 057001
 \bibitem{fang} L. Fang et. al., Phys Rev. B 80 (2009) 140508(R)
\bibitem{pena} J. P. Pe\~na  \textsl{et. al.}, Physica C 531 (2016) 30–38 
\bibitem{myi} M. Yi, \textsl{et. al.}, Phys. Rev. B 80 (2009) 174510
\bibitem{marsik2013} P. Marsik, C. N. Wang,  M. Rossle,  M. Yazdi-Rizi, R. Schuster,  K. W. Kim,  A. Dubroka,  D. Munzar, 
T. Wolf,  X. H. Chen,  and C. Bernhard, Phys. Rev. B 88 (2013) 180508(R) 
\bibitem{yin2} Z. P. Yin, K. Haule, G. Kotliar, Nat.  Phys 7 (2011) 294
\bibitem{gliu2009} Guodong Liu, \textsl{et. al.}, Phys. Rev. B 80 (2009) 134519 
\bibitem{shimojima2010} T. Shimojima,  \textsl{et. al.} Phys. Rev. Lett. 104 (2010) 057002 
\bibitem{zhao} J. Zhao \textsl{et. al.}, Nat. Phys. 5 (2009) 555
  \bibitem{liu} M. Liu \textsl{et. al.}, Nat.  Phys. 8 (2012) 376
\bibitem{yin2011} Z. P. Yin, K. Haule, and G. Kotliar, arXiv:1104.3454v1 (2011) 
\bibitem{yang2009} L. X. Yang, \textsl{et. al.}  Phys. Rev. Lett. 102 (2009) 107002
\bibitem{eom} M. J. Eom,  S. W. Na,  C. Hoch,  R. K. Kremer,  and J. S. Kim, Phys. Rev. B 85, (2012) 024536 
 \bibitem{garitezi} T. M. Garitezi \textsl{et. al.}, Braz J Phys 43 (2013) 223–229
 \bibitem{ychen} Y Chen, Lu X., Wang M., Luo H., and Li S., Supercond. Sci. Technol., 24 (2011) 065004
 \bibitem{canfield1} P. C. Canfield, S. L. Bud’ko, Ni Ni, J. Q. Yan, and A. Kracher. Phys. Rev. B, 80 (2009) 060501
 \bibitem{canfield2} C. Canfield and Sergey L. Bud’ko., Annu. Rev. Condens. Matter Phys., 1 (2010) 27–50
 
\bibitem{A-mackintosh} A. R. Mackintosh, Phys. Rev. Lett. 9 (1962) 90
\bibitem{B-elliott} R. J. Elliott and F. A. Wedgwood, Proc. Phys. Soc. (London) 81 (1963) 846
\bibitem{C-miwa} H. Miwa, Progr. Theoret. Phys. (Kyoto) 29 (1963) 477
\bibitem{D-arajs} S. Arajs, R. V. Colvin and M. J. Marcinkowski, J. Less Common Metals 4, (1962) 46
\bibitem{E-meaden}  G. T. Meaden and P. Pelloux-Gervais, Cryogenics 5 (1966) 227
\bibitem{F-legvold} S. Legvold in “Magnetic Properties of Rare Earth Metals”, ed. R.J. Elliott, Plenum Press (London) 1972, p. 335.
\bibitem{chu2009} Jiun-Haw Chu, J. G. Analytis, C. Kucharczyk, and I. R. Fisher, Phys. Rev. B 79 (2009) 014506

 \bibitem{rosa} F. S. Rosa,\textsl{et. al.}, Scientific Reports  4 (2014) 6252
 \bibitem{garitezi2} T. M. Garitezi \textsl{et. al.}, J. Appl. Phys. 115 (2014) D711
 \bibitem{ziman} J. M. Ziman, Principles of the Theory of Solids, 2nd Edition, Cambridge University Press, 1972.
\bibitem{olariu} A. Olariu, F. Rullier-Albenque, D. Colson, A.  Forget,  Phys. Rev. B 83 (2011) 054518
\bibitem{dhaka2011} R. S. Dhaka, Chang Liu, R. M. Fernandes, Rui Jiang, C. P. Strehlow, Takeshi Kondo, A. Thaler, Jorg Schmalian,
S. L. Budko, P. C. Canfield, and Adam Kaminski, Phys. Rev. Lett. 107 (2011) 267002 
\bibitem{liu2010} Chang Liu, \textsl{et. al.}, Nat. Phys. 6 (2010) 419
\bibitem{chaloupka} J. Chaloupka and G. Khaliullin, PRL 110 (2013) 207205
\bibitem{brouet2013} V. Brouet, Ping-Hui Lin, Y. Texier, 1 J. Bobroff, A. Taleb-Ibrahimi, P. Le Fevre, 2 F. Bertran,  M. Casula, 
P. Werner, S. Biermann, F. Rullier-Albenque,  A. Forget, and D. Colso, Phys. Rev. Lett. 110 (2013) 167002 
\bibitem{deok2009} Eun Deok Mun, Sergey L. Budko, Ni Ni, Alex N. Thaler, and Paul C. Canfield, Phys. Rev. B 80 (2009) 054517
 \bibitem{nagaosa} N. Nagaosa, J. Sinova, S. Onoda, A. H. MacDonald, N. P. Ong, Rev.  Mod. Phys. 82 (2010) 1539
 \bibitem{sinitsyn} N. A. Sinitsyn, J. Phys. Condens. Matter. 20 (2008) 023201
  \bibitem{117dai} H. Gretarsson,\textsl{et. al.}, Phys. Rev. B 84 (2011) 100509
\bibitem{fernandes2011} Rafael M. Fernandes,  Elihu Abrahams, and Jorg Schmalian, Phys. Rev. Lett. 107 (2011) 217002  
\bibitem{wang2015} Y. Wang, Maria N. Gastiasoro,  Brian M. Andersen, M. Tomic, Harald O. Jeschke, Roser Valentí, Indranil Paul, 
and P. J. Hirschfeld, Phys. Rev. Lett. 114 (2015) 097003 
\bibitem{ishida} S. Ishida \textsl{et. al.}, Phys. Rev. B 84 (2011) 184514 
\bibitem{dhaka2013} R. S. Dhaka,  S. E. Hahn, E. Razzoli, Rui Jiang,  M. Shi,  B. N. Harmon,  A. Thaler,  S. L. Budko, 
P. C. Canfield,  and Adam Kaminski, Phys. Rev. Lett. 110, 067002 (2013)



\end{thebibliography}
\end{document}